\documentclass[reprint,aps,prb,showpacs,superscriptaddress,english]{revtex4-2}

\usepackage[utf8]{inputenc}
\usepackage{hyperref}

\usepackage{natbib}
\usepackage{comment}

\usepackage{graphicx,color}
\usepackage{amssymb,amsmath,amsfonts,mathrsfs,fancyhdr}

\usepackage[caption=false,subrefformat=parens,labelformat=empty]{subfig}

\graphicspath{{./figures/}}

\begin{document}

\title{Wetting critical behavior in the quantum Ising model within the framework\\
of Lindblad dissipative dynamics}
\author{Claudia Artiaco}
\email{artiaco@kth.se}
\affiliation{Department of Physics, KTH Royal Institute of Technology, Stockholm 106 91, Sweden}

\author{Andrea Nava}
\affiliation{Dipartimento di Fisica, Università della Calabria, Arcavacata di Rende I-87036, Cosenza, Italy}
\affiliation{INFN - Gruppo collegato di Cosenza, Arcavacata di Rende I-87036, Cosenza, Italy}

\author{Michele Fabrizio}
\affiliation{Scuola Internazionale Superiore di Studi Avanzati (SISSA), via Bonomea 265, 34136, Trieste, Italy}

\begin{abstract}
We investigate the critical behavior, both in space and time, of the wetting interface within the coexistence region around the first-order phase transition of a fully-connected quantum Ising model in slab geometry. For that, we employ the Lindblad master equation formalism in which temperature is inherited by the coupling to a dissipative bath, rather than being a functional parameter as in the conventional Cahn's free energy. Lindblad's approach gives not only access to the dissipative dynamics and steady-state configuration of the quantum wetting interface throughout the whole phase diagram but also shows that the wetting critical behavior can be successfully exploited to characterize the phase diagram as an alternative to the direct evaluation of the free energies of the competing phases. 
\end{abstract}

\maketitle

\section{Introduction}

\label{sec:introduction}

Wetting is an interfacial phenomenon that concerns the ability of liquids to maintain contact with solid, or liquid substrates. Specifically, the study of wetting focuses on understanding the relationship between bulk phase transitions and surfaces. Indeed, a rich variety of phase transitions occur when bulk and surface degrees of freedom are coupled~\cite{cahn1977critical,ebner1977new,lipowski1982critical,Brezin&Halperin1983-2,Coleman-book,de1985wetting,fisher1985wetting,dietrich1988wetting,dietrich1990critical,indekeu1995introduction,telo2003theory}. 
Clearly, this problem is extremely vast and rich. Wetting phenomena have been investigated in a variety of systems ranging from classical ones, such as in liquid-vapor phase transitions or binary liquid mixtures of linear alkanes and methanol, to polymeric mixtures, superfluid $^4$He on thin cesium substrates, liquid $^3$He on superfluid $^4$He, dilute ultra-cold gases undergoing Bose-Einstein condensation, and many others~\cite{cheng1991helium,nacher1991experimental,rutledge1992prewetting,cheng1993novel,taborek1993tuning,kellay1993prewetting,alles1994wetting,indekeu2010wetting}.

In this study, we focus on the dynamics and the equilibrium configuration of the wetting layer that can form within the coexistence region accompanying a quantum first-order phase transition~\cite{bonn2009wetting,rath2011quantum,jakubczyk2012quantum,kaplan2013review,del2016nonequilibrium}. Several attempts to disclose the wetting phenomenon in the quantum realm have relied on the quantum-classical mapping, i.e., on the idea that the properties of $d$-dimensional quantum systems at zero temperature across a phase transition correspond to those of classical systems in higher dimensions~\cite{hertz2018quantum}. Adopting a simple fully connected quantum spin Ising model, the authors of Ref.~\cite{del2016nonequilibrium} observed that the critical properties of wetting in the quantum case indeed correspond to the classical ones in higher dimensions, specifically $d+1$ in that mean-field model; however, they found that the singular behavior of quantum fluctuations is different from that of classical fluctuations at finite temperatures. Commonly, the wetting phenomenon, even in the quantum regime (see, e.g., Ref.~\cite{jakubczyk2015quantum}), has been described within the Landau-Ginzburg framework, as the Cahn's free-energy functional~\cite{cahn1977critical}. There, temperature simply enters as a parameter of the functional that controls the relative depths of the two potential wells, whose crossing defines the first-order bulk phase transition. In this work, instead, we investigate whether the wetting phenomenon can be accessed by the dissipative quantum dynamics in the presence of a thermal bath; dynamics that we approximated via a Lindblad equation (LE) of motion for the density matrix of an exactly-solvable toy model. In the Lindblad approach, the temperature is provided by the bath, and thus it is not a parameter of the system quantum Hamiltonian. We show that our approach is able to recover the conventional wetting critical phenomenon for short-range interactions~\cite{lipowski1982critical}.

Specifically, in this article we consider a slab geometry constituted by $L$ layers, which is a discrete version of the model of Ref.~\cite{del2016nonequilibrium}; each layer is modeled by a quantum Ising model with $N$ fully-connected sites. In the thermodynamic limit, $N\to\infty$,  the mean-field approximation becomes exact, and the equilibrium state of the single-layer system can be found by solving a set of self-consistency equations. The single-layer model possesses already a non-trivial phase diagram that, depending on the form of the spin-spin interactions, can display first- or second-order quantum and thermal phase transitions. Thanks to its simplicity, such a model has been widely employed in the past~\cite{wang1989first,sciolla2011dynamical,bapst2012quantum}, for instance, to study the relaxation dynamics towards equilibrium in the presence of dissipation~\cite{nava2019lindblad,nava2022lindblad,Pei_2021}. We define the single-layer Hamiltonian such that it may undergo a first-order phase transition, and set its parameters so that the single layer is in the coexistence region. In addition, we couple each layer to its nearest neighbor layers, as well as to a dissipative bath by means of the Lindblad master equation (LE)~\cite{wilde_book,breuer2007theory,manzano2020lindblad}. The LE is the most general Markovian, time-local generator for a system density matrix; it is among the most popular master equations, and it has been employed in various and different contexts~\cite{prosen2011open,olmos2012facilitated,brun2000continuous,kraus2008preparation,artiaco2021signatures,nava_2022_ssh}. 

In order to study wetting, which is an inhomogeneous phenomenon, we fix inhomogeneous boundary conditions, i.e., the first (resp.\ last) layer is kept fixed in the ordered (resp.\ disordered) phase of the coexisting region of the phase diagram. Instead, we assume a Lindblad dissipative dynamics for each layer in the bulk, with jump operators defined through the instantaneous mean-field Hamiltonian. This allows us to numerically investigate arbitrarily large system sizes, and to explore the wetting phenomenon without resorting to Cahn's free energy functionals since the temperature of the system is fixed by the bath. We show that, within our model and the LE scheme, the wetting phenomenon spontaneously emerges during the quantum dissipative dynamics; we are able to uncover details of the static and dynamic properties of the wetting interface as a function of the Hamiltonian parameters and the bath temperature. Our results are in agreement with the conventional wetting critical phenomenon for short-range interactions~\cite{lipowski1982critical}. Even though the mean-field nature of our model does not allow studying corrugated configurations of the interface~\cite{rath2011quantum}, our analysis yields a novel result not predicted by Landau-Ginzburg approaches. Indeed, if one unphysically discards longitudinal fluctuations in Cahn's free-energy functional, thus the contribution of capillary waves, and only considers the dynamics of the interface center of mass, the wetting critical phenomenon is not accompanied by any critical behavior of the relaxation time, as can be inferred by Refs.~\cite{Brezin&Halperin1983-2,Coleman-book}. Unlike in Cahn's approach, we show that the Lindblad dynamics of our toy model does predict a critical relaxation that, however, in real systems would be hidden by the slower relaxation due to the critical capillary waves.

The mean-field model introduced in this paper constitutes a promising tool to address the role of quantum fluctuations on wetting phenomena and can be potentially adapted to investigate other complex many-body phenomena in quantum dissipative systems by changing, for instance, the boundary conditions. Recently, appropriate variants of the toy model proposed in this paper have been successfully applied to several contexts, from the Mpemba effect to selective cooling \cite{fabrizio2018selective,nava2019lindblad,nava2022lindblad}.

The paper is organized as follows. In Section~\ref{sec:oneLayer} we present the model Hamiltonian of the single-layer, fully connected quantum Ising model and review its dissipative dynamics yielded by the Lindblad equation. In Section~\ref{sec:multiLayer}, we extend the formalism discussed in the former Section to a multi-layer system in which multiple copies of the single-layer system are connected one to the other to form a slab of length $L$. Each bulk layer is coupled to a dissipative bath while the states of the first and last layers are kept fixed. Section~\ref{sec:results} is devoted to the discussion of the relaxation and equilibrium properties of the multi-layer system. In particular, we analyze the behavior of the layer-resolved equilibrium energy, order parameter, and relaxation time. Finally, in Section \ref{sec:conclusion} we summarize our results and discuss possible future directions of our work.

\section{Single-layer system}

\label{sec:oneLayer}

In this Section, we briefly mention the properties of the single layer when it is decoupled from all the others. In particular, we discuss its phase diagram and show how we construct the Lindblad jump operators to describe its relaxation dynamics.

\subsection{The quantum spin model for the single-layer system}

We model each layer as a quantum Ising model on an $N$-site fully connected graph, described by the general Hamiltonian \cite{wang1989first,sciolla2011dynamical,bapst2012quantum}
\begin{equation}
    \label{eq:generic-H}
    H = -h_{x} \sum_{i}\sigma^{x}_{i}-N\sum_{n=2}^{m}J_{n}\Bigg( \frac{1}{N}\sum_{i}\sigma^{z}_{i}\Bigg)^{n}
    \end{equation}
where $m \geq 2$ is an integer number, $\sigma_i^\alpha$, with $\alpha = x, y, z$, are the Pauli matrices on site $i = 1, \dots, N$, $h_x$ is the transverse magnetic field, and $J_n$ are the $n$-spin exchange constants. In the following, we concentrate on the case $J_2 \neq 0, J_4 \neq 0$, and $J_{n\neq2,4}=0$, for which the model~\eqref{eq:generic-H} undergoes a first-order phase transition \cite{teng2006phase,del2016nonequilibrium}: increasing either the temperature $T$ or the transverse field $h_x$, the system goes from an ordered, ferromagnetic phase (F) to a disordered, paramagnetic one (P). 

\begin{figure}[]
    \centering
    \includegraphics[scale=0.6]{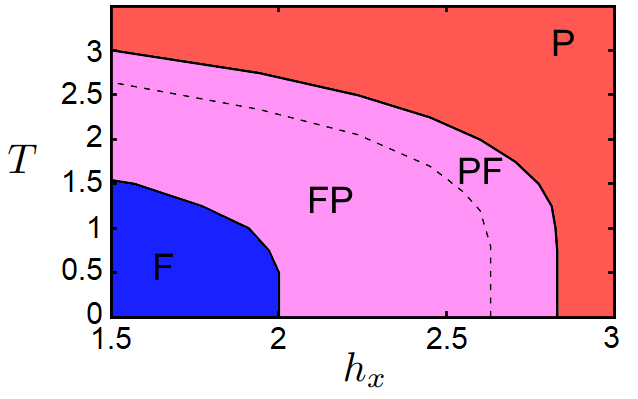}
    \caption{Phase diagram of the single-layer model~\eqref{eq:hamiltonian-oneLayer}. In the region labeled as F, there is only a ferromagnetic free-energy minimum, and the $Z_{2}$ symmetry is broken. Conversely, in region P there is only a paramagnetic minimum. FP and PF regions present three distinct minima: two are ferromagnetic, and one is paramagnetic. In FP, the ferromagnetic minima are absolute minima; in PF, the paramagnetic minimum is the absolute minimum. The F-P phase transition occurs along the dashed line separating FP from PF, and it takes place when the free energies cross. The solid lines between F and FP, and between P and PF, are spinodal lines where an additional metastable free energy minimum appears beside the stable one. Color online.} 
    \label{fig:phaseDiagram}
\end{figure}

In the following, we will express all energies in units of $J_2=J_4=1$. Thus, our single-layer system Hamiltonian reads
\begin{equation}
    \label{eq:hamiltonian-oneLayer}
    H = -h_{x} \sum_{i}\,\sigma^{x}_{i}-\frac{1}{N}\Bigg( \sum_{i}\sigma^{z}_{i}\Bigg)^{2} - \frac{1}{N^3}\Bigg(\sum_{i}\sigma^{z}_{i}\Bigg)^{4}. 
\end{equation}
The phase diagram of the model~\eqref{eq:hamiltonian-oneLayer} has been already studied in the past \cite{teng2006phase,del2016nonequilibrium,nava2019lindblad}; it is illustrated in Fig.~\ref{fig:phaseDiagram}.
    
Thanks to the full connectivity of  Hamiltonian~\eqref{eq:hamiltonian-oneLayer}, mean-field approximation becomes exact in the thermodynamic limit $N\to \infty$, as ensured by the vanishing of the covariance
\begin{equation}
    \langle \sigma_i^\alpha \sigma_j^\beta \rangle - \langle \sigma_i^\alpha \rangle \langle \sigma_j^\beta \rangle \to \frac{1}{N} \xrightarrow[N \to \infty]{} 0\,.
\end{equation}
It follows that the equilibrium Boltzmann distribution is given by
\begin{equation}
    \label{eq:rho_factorized}
    \rho \xrightarrow[N \to \infty]{} \prod_i\, \rho_i = \prod_i \;\frac{e^{-\beta H_i}}{\;\text{Tr}\Big(e^{-\beta H_i}\Big)\;}
\end{equation}
where $\rho_i$ is a positive definite $4 \times 4$ matrix with unit trace, and the single site Hamiltonian $H_i$ reads
\begin{equation}
    \label{eq:Hi-oneLayer}
    H_{i} = -h_{x}\,\sigma^{x}_{i} - h_{z}(\mathbf{m})\,\sigma^{z}_{i},
\end{equation}
with
\begin{equation}
    \label{eq:hz-oneLayer}
    h_{z}(\mathbf{m}) = 2 m_{z} + 4 m_{z}^{3}.
\end{equation}
$\mathbf{m}=(m_x,m_y,m_z)$ indicates the Bloch vector, with components 
\begin{equation}
    m_{\alpha} := \frac{1}{N} \sum_{i}\, \langle\, \sigma_{i}^{\alpha}\, \rangle.
\end{equation}
Notice that the Bloch vector verifies $|\mathbf{m}| \leq 1$, where the equality is fulfilled only by pure states, and that the single-site density matrix can be written as $\rho_i = \frac{1}{2} (1 + \boldsymbol{m} \cdot \boldsymbol{\sigma}_i)$. Thanks to the exact validity of the mean-field approximation, we can consider a single spin at a time and drop the index $i$.    

It is easy to show that, when the system is at equilibrium with a bath at temperature $T$, its state is determined by a set of self-consistency equations:
\begin{equation}
    \label{eq:self-cons}
        \mathbf{m} = \tanh{\beta h(\mathbf{m})}\,\big(\cos\theta(\mathbf{m})\,,\,0\,,\,\sin\theta(\mathbf{m})\big)
\end{equation}
where
\begin{align}
    \label{eq:self-cons2}
    \tan \theta(\mathbf{m}) = \frac{\;h_z(\mathbf{m})\;}{h_x}\,,\\
    \label{eq:self-cons3}
    h(\mathbf{m}) = \sqrt{\,h_x^2+h_z(\mathbf{m})^2\;}\,.
\end{align}

\subsection{Lindblad master equation for the single layer}

The Lindblad master equation (LE) is the most general Markovian master equation. It can be derived starting from a microscopic Hamiltonian describing the interaction between the system (S) and the bath (B)
\begin{equation}
\label{eq:microscopic_H}
    H=H_S+H_B+\alpha H_{int}.
\end{equation}
$H_S$ and $H_B$ denote the system and the bath Hamiltonian, respectively, $H_{int}$ the interaction Hamiltonian, and $\alpha$ the system-bath coupling strength. Under the Markov-Born approximation, i.e., assuming a weak coupling between the system and the bath, $\alpha \ll 1$, and assuming that the bath relaxation time is much smaller than the characteristic timescales of the system, the system-bath density matrix is factorized at any time as
\begin{equation}
    \rho_{S+B}(t) \simeq \rho_S(t) \otimes \rho_B(0)\,, \quad \forall \, t.
\end{equation}
Tracing out the bath degrees of freedom, it is possible to derive an effective equation of motion for the reduced density matrix of the system only that, in the standard Lindblad form, is given by
\begin{multline}
\label{eq:generalLE}
    \dot{\rho_S}(t) = -i\Big[H_S,\rho_S(t)\Big] \\
    + \sum_\lambda\, \bigg[ \gamma_\lambda \Big(2 L_\lambda \rho_S(t)L^\dagger_\lambda - \{L^\dagger_\lambda L_\lambda,\rho_S(t)\}\Big) \\
    + \bar{\gamma}_\lambda \Big(2 L^\dagger_\lambda \rho_S(t) L_\lambda - \{L_\lambda L^\dagger_\lambda,\rho_S(t)\}
    \Big) \bigg]\,,
\end{multline}
with $\lambda = 1,\dots,N^2-1$ where $N$ is the dimension of the Hilbert space of the system~\cite{breuer2007theory}. 
The first term on the r.h.s.\ is the Liouvillian which describes the coherent evolution of the system. The second and third lines constitute the Lindbladian term which describes the incoherent evolution due to the interaction between the system and the bath. The $L_{\lambda}$'s are the so-called “jump operators" that model the stochastic interaction between the system and the bath. The coupling strength $\gamma_{\lambda}$ and $\bar{\gamma}_{\lambda}$ are proportional to $\alpha^2$; their explicit form strongly depends on the microscopic description of the environment through the bath correlation functions. It is worth noticing that, while the LE is in general a first-order linear differential equation for the reduced density matrix, in the presence of a mean-field model (as in this paper) the reduced density matrix becomes non-linear giving rise to a non-trivial behavior.

The LE is the most general dynamical Markovian map that is trace preserving and completely positive (CPT or Kraus map), i.e., it maps density matrices into density matrices. In the following, as we are dealing with a mean-field toy model and in order to remain as general as possible, we introduce the LE within the latter perspective without restoring to any particular microscopic model for the description of the bath (see Refs.~\cite{nathan2020universal,maimbourg2021bath} for two recent microscopic derivations of the LE; in contrast to the quantum optical derivation~\cite{breuer2007theory}, in Refs.~\cite{nathan2020universal,maimbourg2021bath} the LE is derived from a microscopic model without performing the rotating-wave approximation so that this derivation can be applied to any Markovian open quantum system).

In order to describe the system-bath interaction we insert in Eq.~\eqref{eq:generalLE} a complete set of jump operators acting on the system degrees of freedom. In particular, we introduce the energy-conserving, pure-dephasing operators
\begin{equation}
\label{eq:LindbladDephasingOperators}
    L_{\lambda(m,m)}=|m\rangle \langle m |, 
\end{equation}
where $|m\rangle$ are the eigenstates of the system Hamiltonian $H_S$, and
dissipative jump operators that produce transitions between two different eigenstates
\begin{equation}
\label{eq:LindbladJumpOperators}
    L_{\lambda(m,n)}=|m\rangle \langle n |, \quad E_n < E_m. 
\end{equation}
At this point, we only require that the Boltzmann distribution is a stationary solution of Eq.~\eqref{eq:generalLE}, i.e., the bath acts as a dissipative bath at fixed inverse temperature $\beta$. It is easy to verify that this implies a detailed balance condition on the system-bath coupling strength in the form: $\gamma_\lambda/\bar{\gamma}_\lambda = e^{- \beta \epsilon_\lambda}$ with $\epsilon_\lambda = E_m - E_n \geq 0$. Thanks to the freedom left by such a condition, in the following we set
\begin{equation}
\label{eq:gammas}
    \bar{\gamma}_{\lambda} =\Gamma f\big(-\beta \epsilon_{\lambda}/2\big)
\end{equation}
with $f(x)$ the Fermi-Dirac distribution function, and $\Gamma$ the overall system-bath coupling strength~\footnote{The specific choice $\bar{\gamma}_{\lambda} =\Gamma f\big(-\beta \epsilon_{\lambda}/2\big)$ has been mainly motivated by numerical convenience: $f\big(-\beta \epsilon_{\lambda}/2\big)$ is of order 1 for the model parameters $(h_x,T)$ employed in this work. However, we have verified that the results do not change if one employs a Bose-Einstein distribution function, despite the fact that the latter diverges in the limit of 0 energy. The reason is that the instantaneous energy differences $\epsilon(t) \geq 2 h_x$, i.e., they are always finite for $h_x$ finite.}. Let us note that the pure dephasing operators $L_{\lambda(m,m)}$ automatically satisfy the detailed balance condition, as they do not imply any energy transfer; thus, they do not spoil the stationary solution given by the Boltzmann distribution. For this reason, we resort to the so-called “relaxation without pure dephasing" condition~\cite{tempel2011relaxation}, assuming $\gamma_{\lambda(m,m)}=\bar{\gamma}_{\lambda(m,m)}=0$. Such a condition is satisfied in a broad class of microscopic models (see Ref~.\cite{may2004charge} for a detailed discussion) and allows us to compare the results in this paper with previous results obtained on the same system \cite{nava2019lindblad,nava2022lindblad}.

Ref.~\cite{nava2019lindblad} extensively discusses many possible ways in which the Lindblad jump operators can be defined to capture the physics of the model~\eqref{eq:hamiltonian-oneLayer}. Here, we consider one choice that, we will argue, allows us to correctly reproduce the dynamics of the wetting interface in the multi-layer system (see Sec.~\ref{sec:results}),  and recover the semiclassical results of Ref.~\cite{del2016nonequilibrium}. 

Starting from a factorized density matrix, the full connectivity of the model~\eqref{eq:hamiltonian-oneLayer} ensures that it remains factorized at any time:
\begin{equation}
    \rho_S(t) \xrightarrow[N \to \infty]{} \prod_i \, \rho_i(t)
\end{equation}
where $\rho_i(t)$ describes the time evolution of the spin $i$, coupled to a bath at temperature $T$ and in the presence of a time-dependent magnetic field given by Eqs.~\eqref{eq:Hi-oneLayer}-\eqref{eq:hz-oneLayer}:
\begin{align}
    \label{eq:instantaneous_field1}
    \mathbf{h}(t) &:= \mathbf{h}(\mathbf{m}(t)) \\
    &:= \Big( h_x\,,\, 0\,,\, 2 m_z(t) + 4 (m_z(t))^3 \Big)
    \label{eq:instantaneous_field2}
\end{align}
with
\begin{equation}
    \mathbf{m}(t) = \frac{1}{N} \sum_i \,\text{Tr}\Big(\rho_S(t)\boldsymbol{\sigma}_i\Big)\,,
\end{equation}
which is self-consistently determined by the system's time evolution. Notice that the mean-field nature of the model is at the origin of the self-consistency of the dissipative dynamics. Hence, we can formally define a time-dependent system Hamiltonian as
\begin{equation}
\label{eq:instH-oneLayer}
    H_t := -\mathbf{h}(t)\cdot \boldsymbol{\sigma} := - \big|\mathbf{h}(t)\big|\, \mathbf{v}_3(t)\cdot \boldsymbol{\sigma},
\end{equation} 
which is just a two-level system Hamiltonian with a time-dependent magnetic field. From Eq.~\eqref{eq:LindbladJumpOperators}, we can write the instantaneous Lindblad jump operators
\begin{align}
\label{eq:jump-oneLayer}
\notag
    L(t) &= | 1 \rangle \langle 0 | = \big(\mathbf{v}_1(t) - i \mathbf{v}_2(t)\big) \cdot \boldsymbol{\sigma}/2 \\
    &:= \mathbf{v}^-(t) \cdot \boldsymbol{\sigma}/2,
\end{align}
and its Hermitian conjugate
\begin{equation}
\label{eq:jump-oneLayer2}
    L^\dagger(t) = | 0 \rangle \langle 1 | := \mathbf{v}^+(t) \cdot \boldsymbol{\sigma}/2,
\end{equation}
where
\begin{align}
    &\mathbf{v}^+(t) \wedge \mathbf{v}^-(t) = 2 \mathbf{v}_3(t),\\
    &\mathbf{v}^+(t) = \big(\mathbf{v}^-(t)\big)^*.
\end{align}
The energy difference between the eigenvalues of the instantaneous Hamiltonian is simply 
$\epsilon(t) = 2 \big| \mathbf{h}(t)\big|$. Thus, $\bar{\gamma}(t)$ and $\gamma(t)$ depend on time. 

In the case $\mathbf{v}_3(t)=(0,0,1)=\hat{z}$, Eqs.~\eqref{eq:jump-oneLayer}--\eqref{eq:jump-oneLayer2} reduce to $\sigma^{\mp}$ up to an arbitrary phase factor corresponding to a free rotation on the $x-y$ plane. In general, due to the mean-field nature of the system Hamiltonian~\eqref{eq:hamiltonian-oneLayer}, the magnetic field acting on the layer changes magnitude and direction over time (see Eqs.~\eqref{eq:instantaneous_field1}-\eqref{eq:instantaneous_field2}); the Lindblad jump operators change accordingly, following the instantaneous magnetic field on each layer, and being the lifter/lowerer operators of the Hamiltonian at each time.

In turn, this yields a LE with time-dependent parameters, where the expectation value of the spin operator is given by
\begin{align} \notag
    &\dot{\mathbf{m}}(t) = \text{Tr} \Big(\dot{\rho}_S(t) \boldsymbol{\sigma}\Big) = - 2 \mathbf{h}(t) \wedge \mathbf{m}(t) \\ \notag
    &- \frac{\gamma(t)}{2}\, \Big[ 4 \bigl( \mathbf{v}_3(t) + \mathbf{m}(t)\bigr) - \mathbf{v}^-(t) \bigl(\mathbf{v}^+(t) \cdot \mathbf{m}(t)\bigr) \\ \notag
    &\qquad\qquad\qquad - \mathbf{v}^+ \bigl( \mathbf{v}^-(t) \cdot \mathbf{m}(t)\bigr) \Big] \\ \notag
    &+\frac{\bar{\gamma}(t)}{2} \Big[ 4 \bigl( \mathbf{v}_3(t) - \mathbf{m}(t)\bigr) + \mathbf{v}^-(t) \bigl(\mathbf{v}^+(t) \cdot \mathbf{m}(t)\bigr) \\
    &\qquad\qquad\qquad + \mathbf{v}^+(t) \bigl( \mathbf{v}^-(t) \cdot \mathbf{m}(t)\bigr) \Big]\,.
    \label{eq:LE-oneLayer}
\end{align}

It is worth noting that, in our approach, the instantaneous Lindblad operators and the corresponding coupling strengths are functions of the density matrix itself due to the mean-field approximation. To guarantee that the Markovian approximation is fulfilled at any time $t$, we require an instantaneous update of the Lindblad jump operators.
In Ref.~\cite{nava2019lindblad}, it has been shown that the Lindblad jump operators~\eqref{eq:jump-oneLayer}--\eqref{eq:jump-oneLayer2} are not able to capture the long-time dynamics of the single-layer system, which remains trapped at all times in the closest stationary state, even if metastable. Indeed, in the presence of a first-order phase transition, the system can exhibit more than one coexisting phase, whether stable or metastable. In the single-layer system considered here, for a LE with dissipative operators~\eqref{eq:jump-oneLayer}--\eqref{eq:jump-oneLayer2}, each phase has a different “basin of attraction” as a function of the initial condition, i.e., the system is attracted by one of the possible minima depending on the basin of attraction to which the initial condition belongs to. In this case, to describe the full dynamics, both at short and long times, and the relaxation to the true equilibrium state, one needs to write the master equation as a sum of competing terms, one for each phase (either stable or metastable). In this way, both supercooling and the Mpemba effect emerge during the dissipative dynamics. Fortunately, we will not need such a complicated master equation to describe the relaxation and equilibrium dynamics of the wetting layer, as will become clear in the next Sections.

\section{Multi-layer system}
\label{sec:multiLayer}

In this Section, we introduce the multi-layer model for quantum wetting that we are going to investigate in the following. We discuss how we introduce the inhomogeneities at the boundaries, and how we couple each layer in the bulk to a bath in order to study the relaxation and equilibrium dynamics of the wetting interface.

\subsection{The quantum spin model for the multi-layer system}

Let us now consider a multi-layer system composed of $L$ layers, where each layer is modeled by the Hamiltonian in Eq.~\eqref{eq:hamiltonian-oneLayer}, and it is coupled to its nearest neighbor layers via quadratic and quartic terms:
\begin{multline} 
    \label{eq:hamiltonian-multiLayer}
    H_{T} = \sum_{\ell=1}^{L} \, H_\ell - \sum_{\ell = 1}^{L-1} \Bigg\{  \frac{\tilde{J}_2}{N}\Bigg(\sum_{i \in \ell}\sigma^{z}_{i}\Bigg) \Bigg( \sum_{i \in \ell+1}\sigma^{z}_{i}\Bigg)\\
     + \frac{\tilde{J}_4}{N^3} \Bigg( \sum_{i \in \ell}\sigma^{z}_{i}\Bigg)^{2} \Bigg(\sum_{i \in \ell+1}\sigma^{z}_{i}\Bigg)^{2} \,\Bigg\}\,,
\end{multline}
where $H_\ell$ is the Hamiltonian \eqref{eq:hamiltonian-oneLayer} for layer $\ell$. In the following, we set $\tilde{J}_{2,4} = J_{2,4} / 2 = 1 / 2$ so that the equilibrium phase diagram of the homogeneous multi-layer model (i.e., when all layers are in the same state) reduces to the one of the single-layer case in Fig.~\ref{fig:phaseDiagram} with the presence of a coexistence region where both the ferromagnetic, F, and paramagnetic, P, phases are minima of the free energy.

In the thermodynamic limit, the mean-field single site Hamiltonian for layer $\ell$, dropping the site index, reads 
\begin{multline}
    \label{eq:Hi-ellLayer}
    H_{*\,\ell} = -h_{x}\,\sigma^{x}_{\ell} - \Big( m_\ell^z + 2 m_\ell^z{^3} \Big)\,\sigma^{z}_{\ell}\\
    - 1/2 \Big( m^{z}_{\ell-1} + m^{z}_{\ell+1} \Big) \sigma^{z}_{\ell}\\
    - \Big( m^{z}_{\ell-1}{^2} + m^{z}_{\ell+1}{^2} \Big)  m^{z}_{\ell} \,\sigma^{z}_{\ell}.
\end{multline}
We see that when the multi-layer system is in the homogeneous case, i.e., all the layers are in the same state, Eq.~\eqref{eq:Hi-ellLayer} reduces to the single-layer Hamiltonian of Eq.~\eqref{eq:Hi-oneLayer}.

\subsection{Lindblad master equation for the multi-layer system}
\label{subsec:lindblad-multiLayer}

\begin{figure}[]
\centering
\includegraphics[scale=0.35]{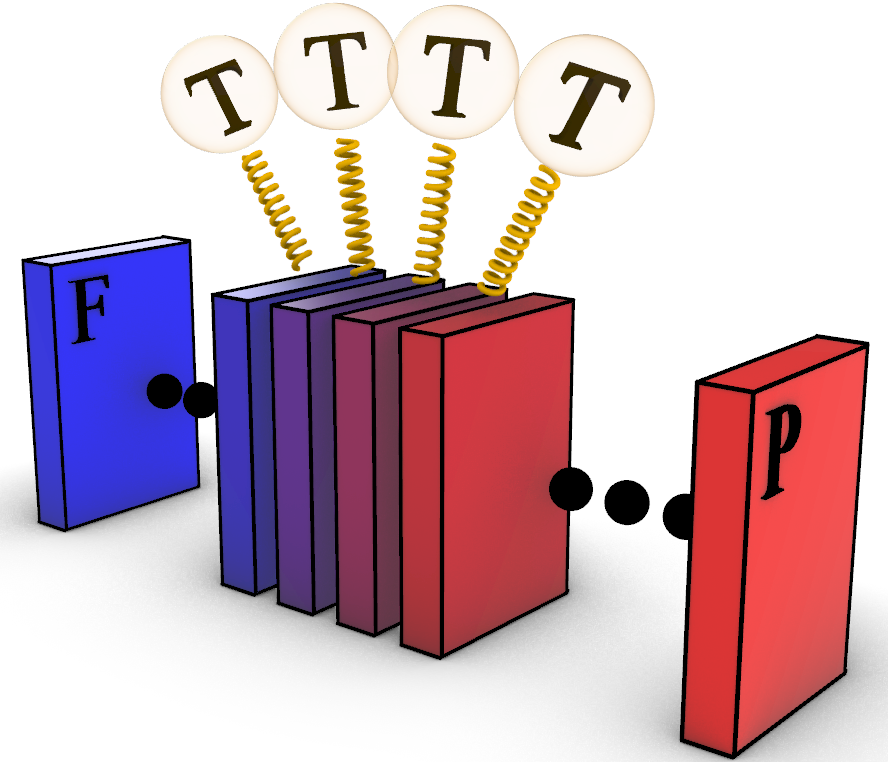}
\caption{Graphical representation of the model. We consider a multi-layer system composed of $L$ layers. The first and the last layers are fixed: they are set to the ferromagnetic (F) and paramagnetic (P) state, respectively. All the other layers are coupled to a heat bath at temperature $T$, as schematized by the springs. Color online.}
\label{fig:multiLayer}
\end{figure}

We wish to study the dynamics of the wetting layer when the single layer is within the coexistence phase illustrated in Fig.~\ref{fig:phaseDiagram}. Within the coexistence phase, the single layer presents both stable and metastable phases. In the FP region, the ferromagnetic minima are stable (i.e., have a lower free-energy), and the paramagnetic is metastable; viceversa in the PF region. To model the presence of the wetting layer, we fix the first and the last layers of the multi-layer system in the ferromagnetic (F) and paramagnetic (P) phase, respectively. All other layers are coupled to a heat bath and are free to evolve in time. The system is depicted in Fig.~\ref{fig:multiLayer}.

Following the same line of reasoning of Sec.~\ref{sec:oneLayer} for the case of a single layer, we can write the dissipative dynamics of the full system considering a time-dependent magnetic field for each layer $\ell$ which is self-consistently determined by the system dynamics.
From Eq.~\eqref{eq:Hi-ellLayer}, the time-dependent magnetic field, accounting for both the intra- and inter-layer interactions is
\begin{align}
    \label{eq:multiLayer-field}
    \notag
    \mathbf{h}_{\ell}(t) &:= \Big| \mathbf{h}_{\ell}\big(\mathbf{m}_{\ell}(t),\mathbf{m}_{\ell-1}(t),\mathbf{m}_{\ell+1}(t)\big) \Big| \\ &\qquad\qquad \mathbf{v}_{\ell}^{3}\big(\mathbf{m}_{\ell}(t),\mathbf{m}_{\ell-1}(t),\mathbf{m}_{\ell+1}(t)\big) \\ \notag
    &:= \biggl[ h_{x}\,,\,0\,,\,m^{z}_{\ell}(t) + 2\,m^{z}_{\ell}(t){^3} \\\notag
    &\qquad\;  + 1/2\, \bigl( m^{z}_{\ell-1}(t) + m^{z}_{\ell+1}(t) \bigr) \\
    &\qquad\; + \bigl( m^{z}_{\ell-1}(t){^2} + (m^{z}_{\ell+1}(t){^2} \bigr)\, m^{z}_{\ell}(t) \biggr]\,.
\end{align}
From Eq.~\eqref{eq:multiLayer-field}, we can define the single-layer time-dependent jump operators, similarly to Eqs.~\eqref{eq:jump-oneLayer}--\eqref{eq:jump-oneLayer2}, the only difference being that both the magnitude and the direction of the time-dependent magnetic field $\mathbf{h}_{\ell}(t)$ vary between layers. In fact, for each layer $\ell$, we obtain a LE for the expectation value of the magnetization similar to Eq.~\eqref{eq:LE-oneLayer}, in which $\mathbf{h}(t), \mathbf{v}_3(t), \mathbf{v}^+(t), \mathbf{v}^-(t)$ depend on $\mathbf{m}_\ell(t), \mathbf{m}_{\ell-1}(t), \mathbf{m}_{\ell+1}(t)$. Notice that the presence of the coupling between layers gives rise to a set of coupled first-order non-linear differential equations that determine the wetting dynamics. Our implementation of the numerical solver for such a set of differential equations is provided at Ref.~\cite{code2022github}.

\section{Results}
\label{sec:results}

In this Section, we discuss the results of the relaxation dynamics and equilibrium configuration of the multi-layer quantum spin model described in Sec.~\ref{sec:multiLayer}, where inhomogeneities are introduced across the boundaries of the system in the presence of a first-order phase transition. In the following, we always set $T$ and $h_x$ within the coexistence region of the single layer model~\eqref{eq:hamiltonian-oneLayer}.
In order to study interface phenomena, we consider a finite length $L$, multi-layer system, as depicted in Fig.~\ref{fig:multiLayer}, with the first and last layers, i.e., the boundaries, fixed in the F and P state respectively. At equilibrium, we expect that when $T$ and $h_x$ are in the FP (PF) phase, the bulk of the system lies in the F (P) phase while a small but finite region, i.e. the wetting region, forms near the last (first) layer of the slab. In the following, we discuss the energy cost and thickness of the wetting region as a function of the temperature and the magnetic field. Moreover, we discuss the dependence of the relaxation time on the bath coupling strength.

\subsection{Energy}
\label{subsec:ene}

Let us start by looking at the relaxation dynamics in the $T \simeq 0.07 $ (i.e., $T \approx 0$) limit: the coexistence region extends from $h_x \simeq 2$ to $h_x \simeq 2.83$, and the critical magnetic field separating the FP and PF phases is $h_c \simeq 2.63113$. At $t<0$, the system is prepared in its equilibrium minimum, i.e., all the layers are in the F (P) phase for $h_x$ lower (higher) than $h_c$. The state of each layer is specified by the Bloch vector $\mathbf{m}_\text{F}$ ($\mathbf{m}_\text{P}$) self-consistently determined by Eqs.~\eqref{eq:self-cons}--\eqref{eq:self-cons3}. At $t=0$ we fix the first and last layers into the F  and P phase, respectively; at the same time, each layer in the bulk is connected with its own bath (see Fig.~\ref{fig:multiLayer}). Suddenly, the Bloch vector of each layer, $\mathbf{m_{\ell}}$, starts to evolve in time while each layer exchanges energy with its neighbors and with the bath in order to reach the new equilibrium configuration that minimizes the energy of the multi-layer system, compatibly with the inhomogeneity introduced via the boundaries.

\begin{figure}[]
\centering
\includegraphics[width=\columnwidth]{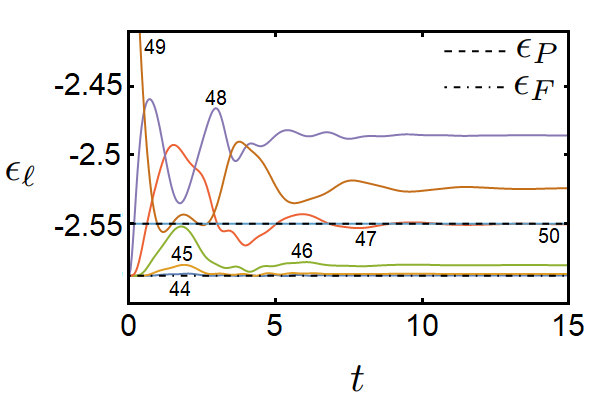}
\includegraphics[width=\columnwidth]{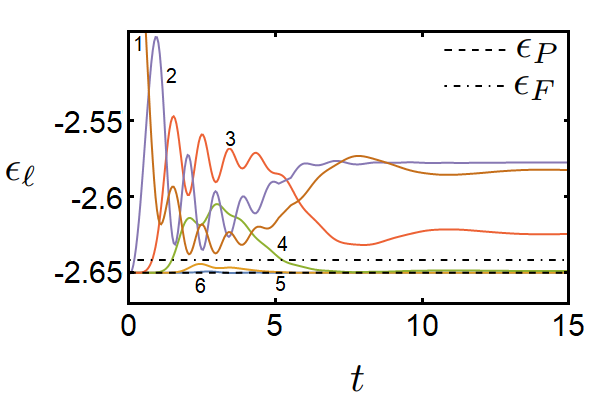}
\caption{Layer-resolved time evolution of the energy for a system of $L=50$ layers for $\Gamma=0.2$, in the FP phase ($h_x=2.55$, top panel), and in the PF phase ($h_x=2.65$, bottom panel). In both cases, the first layer is constrained into the F phase and the last layer into the P phase. The dot-dashed line is the energy of the homogeneous F phase ($\epsilon_\text{F}$), and the dashed line is the energy of the homogeneous P phase ($\epsilon_\text{P}$). Only the layers closer to the corresponding metastable boundary are explicitly shown as the energy of the layers in the bulk overlaps with the energy of the stable minima, i.e. $\min[\epsilon_\text{F},\epsilon_\text{P}]$. The label $\ell$ of each layer is shown near each curve. Color online.}
\label{fig:energy_evo}
\end{figure}

In Fig.~\ref{fig:energy_evo} we plot the time evolution of the single-layer energy, defined as
    \begin{multline}
    \label{eq:en-ellLayer}
       \epsilon_{\ell} = -h_{x}m^{x}_{\ell} - \frac{1}{2} \,m_\ell^z{^2} - \frac{1}{2}\, m_\ell^z{^4}\\
      -\frac{1}{4} \,\bigl( m^{z}_{\ell-1} + m^{z}_{\ell+1} \bigr)\, m_\ell^z\\
      - \frac{1}{4}\, \bigl( (m^{z}_{\ell-1})^2- (m^{z}_{\ell+1})^2 \bigr) \,m_\ell^z{^2},
    \end{multline}
obtained numerically by integrating the nonlinear LE given by Eqs.~\eqref{eq:LE-oneLayer} and \eqref{eq:multiLayer-field}. We consider a bath coupling strength $\Gamma=0.2$ and magnetic field along $x$ below, $h_x=2.55$ (top panel), and above, $h_x=2.65$ (bottom panel), the critical value $h_c$. From now on, we refer with $\epsilon_\text{F}$ ($\epsilon_\text{P}$) to the single-layer energy of a homogeneous system with magnetization vector $\mathbf{m}_\text{F}$ ($\mathbf{m}_\text{P}$), i.e., $\epsilon_\text{F}\equiv\epsilon_{\ell}(\mathbf{m_{\ell}}=\mathbf{m}_\text{F}\ \forall \ell)$ ($\epsilon_\text{P}\equiv\epsilon_{\ell}(\mathbf{m_{\ell}}=\mathbf{m}_\text{P}\ \forall \ell)$). 

Notice that the single-layer energy has an intra-layer contribution plus an inter-layer term. At $t=0$, when the inhomogeneities at the boundaries are created, all the layers deep in the bulk of the system have the same energy as in the stable homogeneous configuration. On the contrary, a large energy contribution emerges from the inter-layer term at the metastable boundary, due to the discontinuity in the magnetization between the boundary layer in the metastable phase ($\ell=L$ for the FP case and $\ell=1$ for the PF one) and its neighboring layer (see the brown curve in both panels of Fig.~\ref{fig:energy_evo}). In the early stage of dynamics, in order to reduce the total energy, the order parameters of the layers around the metastable boundary start to rearrange assuming intermediate values between $\mathbf{m}_\text{F}$ and $\mathbf{m}_\text{P}$ in order to make the discontinuity smoother. Doing so, while the intra-layer energy contribution increases, the inter-layer energy of the boundary drastically decreases driving the system, at large $t$, into a new inhomogeneous equilibrium configuration with the formation of a wetting interface.

\begin{figure}[]
\centering
\includegraphics[width=\columnwidth]{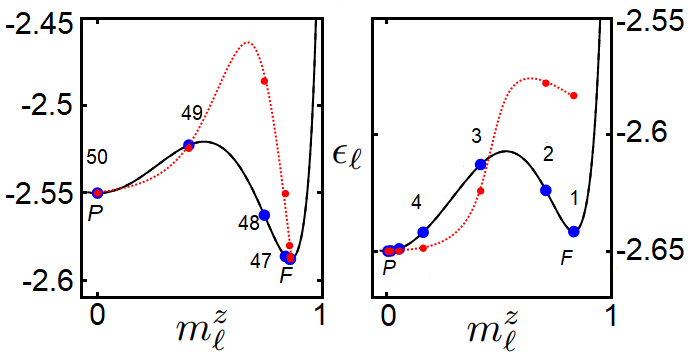}
\caption{Black line: energy per layer, $\epsilon_\ell$, of the homogeneous system as a function of the order parameter for $h_x=2.55$ (FP, left panel) and $h_x=2.66$ (PF, right panel). Blue dots: energy per layer in a homogeneous system at fixed magnetization $m^z_\ell$. Red dots: equilibrium energy of the layers closer to the metastable boundary in the presence of the wetting interface (i.e., corresponding to the equilibrium values in Fig.~\ref{fig:energy_evo}). The red dotted line is a guide for the eye, representing the (pseudo)potential for the inhomogeneous system. Color online.}
\label{fig:pseudo}
\end{figure}

At equilibrium, moving along the wetting interface, from the metastable boundary towards the bulk, the single-layer energy $\epsilon_\ell$ is a non-monotonous function of the distance from the boundary. Plotting the equilibrium values of $\epsilon_{\ell}$ as a function of the magnetization $m^z_\ell$ (see Fig.~\ref{fig:energy_evo}), we observe that the energy increases and then decreases until it reaches the stable value $\min[\epsilon_\text{F}, \epsilon_\text{P}]$ inside the bulk as if it were virtually climbing up the (pseudo)potential barrier that separates the metastable and stable phases at the boundaries. This interpretation becomes clearer if we refer to the energy that each layer would have in the homogeneous case at fixed  $m^z_\ell$, which we plot in Fig.~\ref{fig:pseudo} (blue dots) together with the equilibrium energies (red dots) of the layers forming the wetting interface, same data of Fig.~\ref{fig:energy_evo}. We note that the blue dots strictly follow the energy landscape, climbing up the potential barrier that separates the metastable minima from the stable one, while the red dots describe a new energy landscape of the inhomogeneous system.

We emphasize that, with our choice of jump operators, albeit the single-layer system would remain forever trapped in the metastable phase~\cite{nava2019lindblad}, we do not observe the presence of any metastable phase in the multi-layer system which always thermalizes to the same state. Indeed, the profile of the wetting region results independent of the initial state and the system-bath coupling strength $\Gamma$. The robustness of the steady state has been tested for different physical initial conditions and appears to be independent of them. 
For instance, if we initialize the bulk layers in the metastable minimum, only the dissipation dynamics and the total relaxation time are affected but \emph{not} the equilibrium configuration, which is always the same state. The same happens if we start with an inhomogeneous configuration where P and F regions alternate. To further validate the absence of any metastable state in the multi-layer model, we have explicitly checked, for a three- and four-slab model, that the self-consistency equations only admit a single solution.

We also mention that, if we start in the FP phase and initialize the bulk with domain walls between F$^+$, i.e., $m^z_\ell >0$,  and F$^-$, i.e., $m^z_\ell <0$, regions, interfaces emerge with $m_{\ell}^z$ switching from positive to negative passing through layers in the disordered phase $m_\text{P}^z$. Such an interface has exactly the same shape as the wetting interface at the boundary between the P and F phases but is now doubled around the central P layer. Although these composite $\text{F}^{\pm}-\text{P}-\text{F}^{\mp}$ domain walls have an energy cost, they can move inside the system and annihilate with each other. This behavior is similar to the formation and propagation of solitons in trans-polyacetylene chains~\cite{heeger1988solitons}. Studying the dynamics of domain walls between two equivalent stable minima in the coexistence region goes beyond the scope of this work and will be discussed in a forthcoming publication. This effect does not emerge in the PF phase where the $\text{F}^{\pm}-\text{F}^{\mp}$ domain walls are destroyed by the bath in favor of the stable ordered phase P.

\subsection{Thickness}
\label{subsec:thick}

The wetting layer extent is expected to increase in size as the critical magnetic field $h_c$ is approached. Indeed, as $h \rightarrow h_c$, one has $\left|\epsilon_\text{F}-\epsilon_\text{P} \right| \rightarrow 0$. It follows that, in order to smooth out the discontinuity, the system prefers to unpin more and more layers at the boundary from the stable phase towards the metastable one. In fact, the nearer we are to the critical magnetic field, the lower the intra-layer energy contribution is, while the inter-layer one becomes dominant. 

\begin{figure}[]
\centering
\includegraphics[width=\columnwidth]{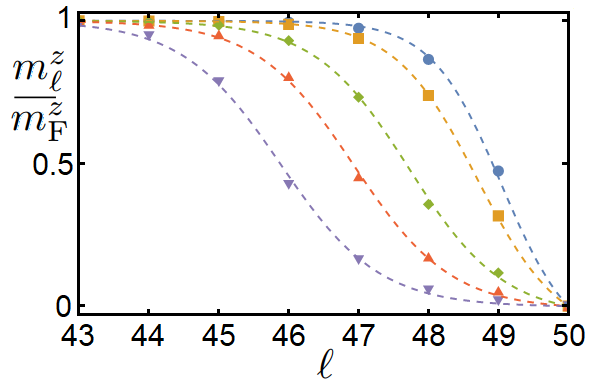}
\includegraphics[width=\columnwidth]{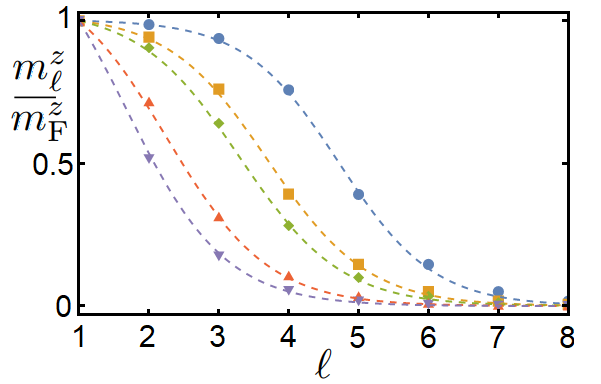}
\caption{Layer-resolved equilibrium value of the order parameter in the FP (top panel) and PF (bottom panel) phase for $\Gamma=0.2$ and different values of the magnetic field along x, $h_x$. Only the eight nearest layers to the metastable boundary are shown. Top panel: $h_x=2.55$ (blue circles), $2.6$ (yellow squares), $2.628$ (green diamonds), $2.6295$ (red triangles), $2.631$ (purple inverted triangles). Bottom panel: $h_x=2.632$ (blue circles), $2.635$ (yellow squares), $2.64$ (green diamonds), $2.7$ (red triangles), $2.8$ (purple inverted triangles). The dashed lines correspond to the fit obtained through Eq.~\eqref{eq:tanh}. Color online.}
\label{fig:thickness}
\end{figure}
In the PF phase ($h > h_c$), we define the amplitude of the wetting interface due to the F phase as
\begin{equation}
   \mathcal{A}_{\text{F}}= \sum_{\ell=1}^L \,\frac{\displaystyle \;m^z_{\ell}\;}{\displaystyle m^z_\text{F}}\;,
\end{equation}
while in the FP phase ($h<h_c$) we define the wetting amplitude due to the P phase as
\begin{equation}
       \mathcal{A}_{\text{P}}=\sum_{\ell=1}^L\, \bigg( 1 - \frac{\displaystyle \;m^z_{\ell}\;}{\displaystyle m^z_\text{F}}
       \bigg) = L-\mathcal{A}_{\text{F}}
       \,.
\end{equation}
By looking at the wetting surface shown in Fig.~\ref{fig:thickness}, we observe that, even for $\left| h_c-h_x \right| \approx 10^{-4}$, only a small finite number of layers around the metastable boundary is involved, so that the equilibrium configuration is not affected by the system length $L$, i.e. $\mathcal{A}_\text{F/P}\ll L$. Clearly, in the case $\mathcal{A}_\text{F/P}\approx L$ the system length must be increased accordingly in order to avoid finite size effects. The data in Fig.~\ref{fig:thickness} can be fitted by a two-parameter function of the form
\begin{equation}
\label{eq:tanh}
   f\left( \alpha, \beta \right)=\frac{\;\tanh{\left( \beta L -\alpha \right)}-\tanh{\left( \beta \ell - \alpha \right)}\;}{\tanh{\left( \beta L -\alpha \right)}-\tanh{\left( \beta - \alpha \right)}}\,,
\end{equation}
similar two the solitons bond-alternation domain walls in polyacetylene~\cite{heeger1988solitons}, or the non-equilibrium stationary state occupation number profile of an interacting fermionic chain~\cite{nava2021lindblad}. This behavior is substantially different from the exponential decay expected for the second-order phase transitions~\cite{borghi2009surface}.

\begin{figure}[]
\centering
\includegraphics[width=\columnwidth]{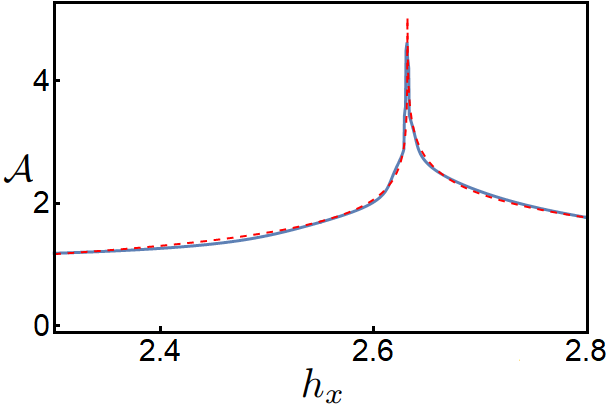}
\includegraphics[width=\columnwidth]{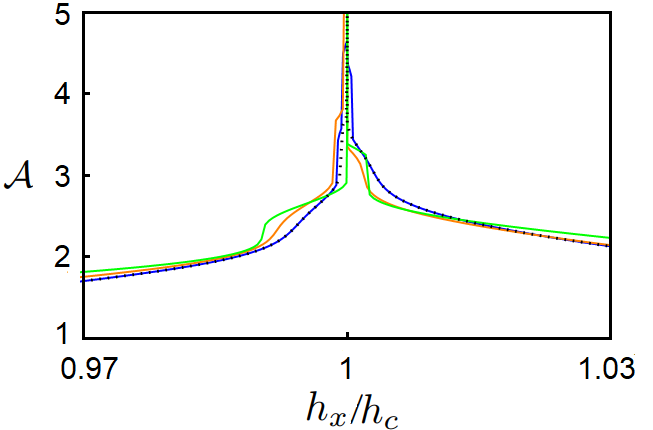}
\caption{Top panel: wetting amplitude $\mathcal{A}=\min[\mathcal{A}_{\text{F}},\mathcal{A}_{\text{P}}]$ of the wetting region as a function of the transverse magnetic field $h_x$ for $T=0.2$ and $\Gamma=0.2$ (blue curve). Logarithmic fit $f(h_x) := a-b \ln{\left| h_x -h_c \right|} $ with $a=0.75$, $b=0.38$ for $h_x<h_c$, and $a=1$, $b=0.43$ for $h_x>h_c$ (red dashed curve).
Bottom panel: wetting amplitude as a function of the transverse magnetic field $h_x/h_c(T)$ for $T=0$ (black dots), $T=0.2$ (blue curve), $T=1.5$ (orange curve), $T=2$ (green curve). All the curves are rescaled with respect to $h_c \left( T \right)$. Color online.}
\label{fig:logarith}
\end{figure}

In the top panel of Fig.~\ref{fig:logarith} we show the wetting amplitude $\mathcal{A}$ as a function of the magnetic field $h_x$ for a system at finite temperature $T=0.2$. As already discussed, the wetting interfaces increase approaching the critical magnetic field and diverge at $h_x=h_c$, where the phase transition between the FP and PF phases takes place. In agreement with the continuum limit discussed in Ref.\onlinecite{del2016nonequilibrium}, we find that the wetting thickness diverges logarithmically as $\mathcal{A}_{\text{F/P}}=a_\text{F/P}-b_\text{F/P} \ln{\left| h_x -h_c \right|}$, similarly as in the classical case at the thermal phase transition~\cite{lipowski1982critical}. 
In the bottom panel of Fig.~\ref{fig:logarith} we compare the wetting amplitude for different temperatures ranging from $T=0$ to $T=2$. The x-axis is rescaled with respect to the corresponding $h_c(T)$. The behavior of $\mathcal{A}$ is similar at all temperatures: it presents a sharp peak in correspondence of the critical value. Increasing $T$, we observe the formation of a “stepped structure". Its origin can be attributed to two main factors. First, our model is discrete; thus, each step in the behavior of $\mathcal{A}(h_x)$ signals that one more layer has become part of the wetting interface. Second, at high temperatures, upon increasing $h_x$ one approaches the critical line in a non-perpendicular way, surfing in its neighborhood, as can be seen from Fig.~\ref{fig:phaseDiagram}.

\begin{figure}[]
\centering
\includegraphics[width=\columnwidth]{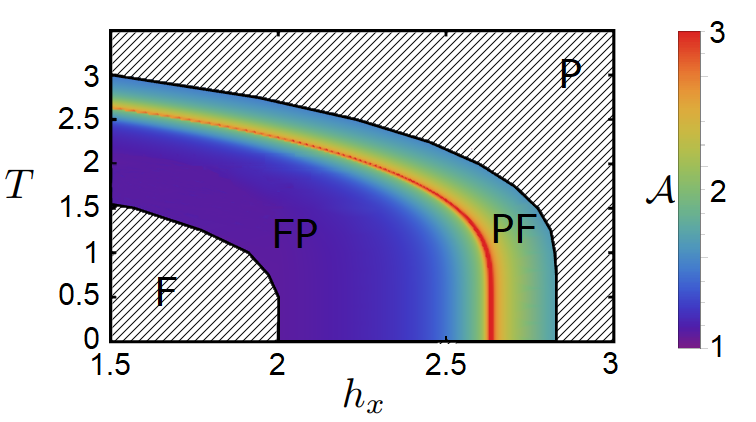}
\caption{$\mathcal{A}=\min[\mathcal{A}_{\text{F}},\mathcal{A}_\text{P}]$ of the wetting layer as a function of the magnetic field and temperature, for $\Gamma=0.2$. Color online.}
\label{fig:phase_area}
\end{figure}

The results of Fig.~\ref{fig:logarith} can be extended at any temperature and any magnetic field within the coexistence region of the homogeneous system, see Fig.~\ref{fig:phaseDiagram}. In particular, the discussions above remain valid by reinterpreting our results in terms of free energy rather than energy at zero temperature~\cite{del2016nonequilibrium}. In Fig.~\ref{fig:phase_area}, we report the wetting amplitude for all the $\left( T, h_x \right)$ values corresponding to the coexistence region of the single-layer phase diagram. 
Just by looking at the divergence of the wetting amplitude $\mathcal{A}$, we are able to recover FP-PF critical line, as can be observed by comparing Fig.~\ref{fig:phase_area} with the single-layer phase diagram in Fig.~\ref{fig:phaseDiagram}. 

This result is nontrivial. In order to access the critical line that separates the two metastable phases in a first-order phase transition, standard approaches require solving self-consistent equations of the form of Eqs.~\eqref{eq:self-cons}--\eqref{eq:self-cons3} and then comparing all the free energy minima at a given value of the parameters (such as temperature and magnetic field). This requires computing, as soon as $T\neq 0$, the entropy of the system, which is often a cumbersome task. In addition, standard LE approaches are only able to recover the F-FP and PF-P critical lines, while failing to recover the FP-PF transition line within the coexistence region~\cite{nava2019lindblad}. On the contrary, in the multi-layer setup discussed in this paper, thanks to the inhomogeneities introduced by the boundary conditions, it is possible to implement the standard self-consistent LE approach to the full phase diagram. Within the LE approach, the only required ingredients are the instantaneous Hamiltonian eigenvectors that define the Lindblad jump operators, see Eq.~\eqref{eq:LE-oneLayer}, at the given bath temperature, see Eq.~\eqref{eq:gammas}. It follows that the full phase diagram at any finite $T$ can be easily derived. 

\begin{figure}[]
\centering
\includegraphics[width=\columnwidth]{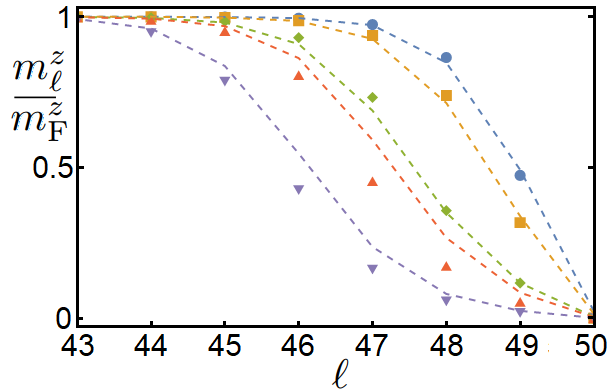}
\includegraphics[width=\columnwidth]{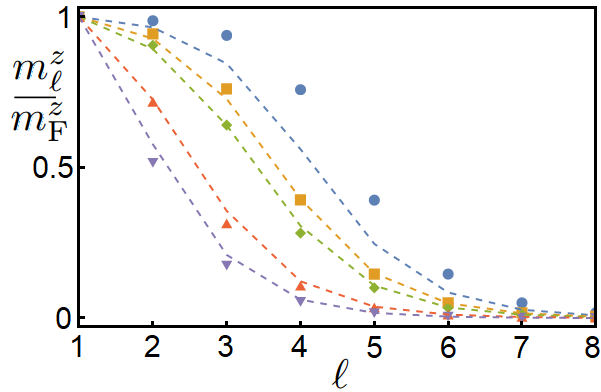}
\caption{Layer-resolved equilibrium value of the order parameter, normalized to the corresponding $m^z_\text{F}$, in the FP (top panel) and PF (bottom panel) phase, for $\Gamma=0.2$ and different values of the transverse magnetic field, $h_x$. The markers represent the discrete model, and the dashed lines the continuum limit within the semiclassical approximation (see main text). Only the eight nearest layers to the metastable boundary are shown. Top panel: $h_x=2.55$ (blue circles), $2.6$ (yellow squares), $2.628$ (green diamonds), $2.6295$ (red triangles), $2.631$ (purple inverted triangles). Bottom panel: $h_x=2.632$ (blue circles), $2.635$ (yellow squares), $2.64$ (green diamonds), $2.7$ (red triangles), $2.8$ (purple inverted triangles). Color online.}
\label{fig:thickness_compare}
\end{figure}

Before moving on to the next Section, let us compare the results obtained within the LE approach for the multi-layer system with the results obtained within the semiclassical analysis on a continuum semi-infinite slab, see Eq.~(35) in Ref.~\cite{del2016nonequilibrium}. In Fig.~\ref{fig:thickness_compare}, we plot, for different values of $h_x$, $m^z_\ell/m^z_\text{F}$ as a function of the layer index $\ell$, obtained within the LE approach (dots), together with the same quantity computed within the continuum limit formula~\cite{del2016nonequilibrium} (dashed lines). We observe a remarkable quantitative agreement between the results of the two approaches as long as $h_x$ is not too close to $h_c$. As $h_x \rightarrow h_c$, the wetting interface of the discrete model is thicker than the continuum one (the leftmost in the top panel, the rightmost in the bottom panel). Such disagreement simply derives from the fact that the interface width is controlled in the continuum limit by the stiffness term, second-order expansion in the interlayer distance of 
the coupling among layers, and that can well change quantitatively the results, but not the critical behavior.

\subsection{Time}
\label{subsec:time}

In this Section, we discuss the behavior of the relaxation time $\tau$ as a function of the magnetic field $h_x$ for different values of the bath coupling strength $\Gamma$ and temperature $T$. The relaxation time $\tau$ measures the time required by the system to reach thermal equilibrium. From the practical point of view, we define $\tau$ in the following way. For each layer, we compute the order parameter $m_z(t)$ as a function of time. The order parameter of each layer tends to a stationary value for $t \to \infty$. We define $\tau$ as the time at which $| m_z(\tau)-m_z(\tau')| < 10^{-10}$, $\forall \; \tau'>\tau$.

\begin{figure}[]
\centering
\includegraphics[width=\columnwidth]{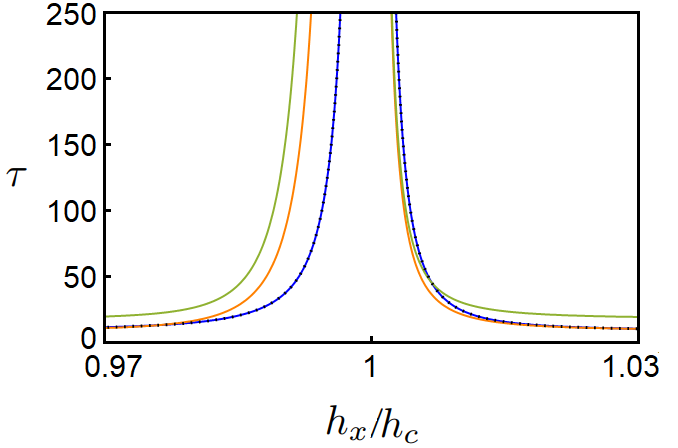}
\includegraphics[width=\columnwidth]{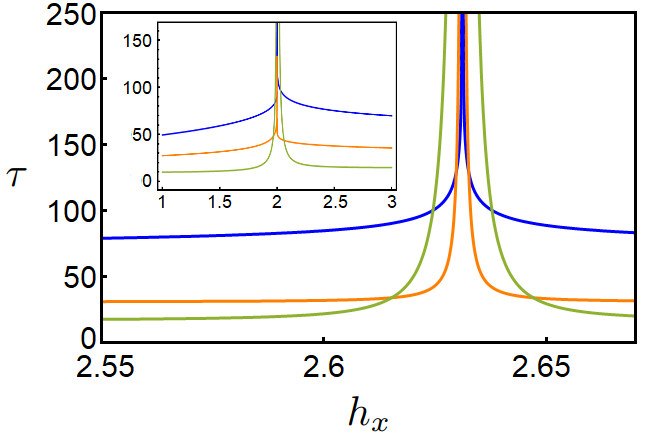}
\caption{Top panel: relaxation time to reach the equilibrium configuration for a system initialized in the F phase, if $h<h_c$, and P phase, if $h>h_c$, as a function of $h_x$, for $\Gamma=0.9$, and $T=0$ (black dots), $T=0.2$ (blue/lower), $T=1.5$ (orange/middle), $T=2$ (green/upper). Bottom panel: as in the top panel, for $T=0.2$, and $\Gamma=0.1$ (blue/upper), $0.2$ (yellow/middle), and $0.5$ (green/lower). The first layer is constrained into the F phase and the last layer into the P phase. Inset: relaxation time for a second-order phase transition, where the critical magnetic field is $h_c=2$. Color online.}
\label{fig:time}
\end{figure}

In the top panel of Fig.~\ref{fig:time}, we show $\tau(h_x/h_c(T))$ for $\Gamma=0.9$ and four different values of the bath temperature, $T=0.0,\,0.2,\,1.5,\,2.0$. While we observe no dependence of $\tau$ on $T$ for $T\lesssim 0.8$, an asymmetric behavior between the ferromagnetic ($h_x<h_c$) and paramagnetic ($h_x>h_c$) phases emerges and becomes more evident as the temperature increases. Such a feature is due to the bending of the critical line of the single-layer phase diagram shown in Fig.~\ref{fig:phaseDiagram}. Indeed, at high temperatures upon changing $h_x$ we do not cross the critical line in a perpendicular way, but surf around it.

In the bottom panel of Fig.~\ref{fig:time}, we show $\tau(h_x/h_c)$ at $T=0.2$ and three different values of the bath strength $\Gamma$. Far from $h_c$, the relaxation time is independent of $h_x$ and reaches a steady value that, as intuitively expected, decreases for increasing bath strength $\Gamma$, i.e., a stronger bath dissipates faster. At any $\Gamma$ we observe a critical slowing down: the relaxation time has a power-law divergence approaching $h_c$, $\tau \sim \left|h_x-h_c \right|^{-\alpha}$, with critical exponent $\alpha$ that is a function of the bath coupling strength $\Gamma$, $\alpha\left(\Gamma \right)$. We find $\alpha(0.2)=1.4$, $\alpha(0.5)=1.8$, and $\alpha(0.9)=2.1$, respectively. A similar dependence of the critical exponent $\alpha$ on $\Gamma$ has been already observed in the literature (see for example Table~II and Fig.~1 in Ref.~\cite{jo2021mean}, and Table~II in Ref.~\cite{jo2021absorbing}) in the case of non-dissipative baths. However, to the best of our knowledge, not much is known about the dependence of the relaxation time on $\Gamma$ within dissipative dynamics.

We observe that the above result escapes a Landau-Ginzburg type of approach once longitudinal fluctuations in Cahn's free-energy functional are discarded. Indeed, in the latter, when the wetting interface is assumed to be infinitely stiff, the wetting critical phenomenon is not accompanied by a critical behavior of the relaxation time. The reason is that the unbinding of the wetting interface is not accompanied by the softening of a continuous spectrum of excitations if one neglects capillary waves~\cite{Brezin&Halperin1983-2,Coleman-book}. Such exclusion seems unrealistic in physical systems but can be legitimately assumed in a model calculation as ours. Unlike in Cahn's approach, the Lindblad dynamics of our toy model does predict a critical relaxation time that unsurprisingly depends on the system-bath coupling. However, since 
$\tau\sim \big| h_x-h_c\big|^{-\alpha}$, with $\alpha > 1$, 
the corresponding relaxation process decay faster than that one provided by the critical behavior of the capillary waves, and which is associated to a smaller exponent, $\alpha_{cw}=1/2$ for short-range interactions~\cite{del2016nonequilibrium}. That is comforting, since it entails a critical dynamics following conventional hyperscaling in physical systems, even in the presence of non-critical dissipative channels that should always exist.

\noindent

\begin{figure}[]
\centering
\includegraphics[width=\columnwidth]{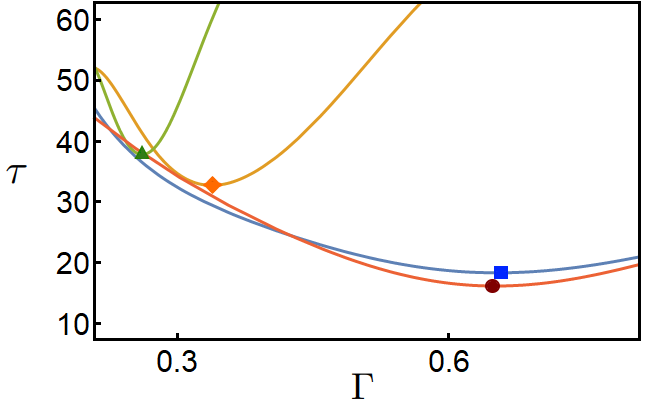}
\caption{Relaxation time to reach the equilibrium configuration for a system initialized in the F phase, if $h<h_c$, and P phase, if $h>h_c$, as a function of $\Gamma$ for $T=0.2$ and different values of the magnetic field in the proximity of $h_c$. The marks represent the optimal working points (minima of the curve) for $h_x=2.60$ (blue square), $2.62$ (yellow rhombus), $2.62$ (green triangle), and $2.66$ (red circle). The first layer is constrained into the F phase and the last layer into the P phase. Color online.}
\label{fig:optimal}
\end{figure}

As a consequence of the dependence of $\alpha$ on $\Gamma$, relaxation time curves for different $\Gamma$ intersect with each other for some value of the magnetic field $h_x$. It follows that for each value of $h_x$ we can define an optimal dissipation strength $\Gamma$ that maximizes the system-bath energy exchange rate. Such a dissipative optimal working point, represented by a minimum of $\tau(\Gamma)$, as shown in Fig.~\ref{fig:optimal} for four different values of the magnetic field, is similar to the non-equilibrium optimal working point that emerges in the non-equilibrium stationary state of systems coupled with two baths \cite{rossini2009negative,rossini2009charge,nava2021lindblad}. Indeed, when a system is coupled to two different baths that induce a particle/energy flow, one observes a change in the monotonicity of the non-equilibrium stationary current as a function of the applied bias, which represents the optimal performance of the bath.

The existence of an optimal working point is a consequence of the presence, in dissipative dynamics, of two timescales: an intrinsic timescale induced by the Hamiltonian of the system which in our case is related to $h_x$, and a dissipative timescale set by the bath strength $\Gamma$. As long as $\Gamma \ll h_x$, increasing $\Gamma$ increases energy exchange and tends to make relaxation faster, i.e., reduces $\tau$. When the two timescales are comparable, we observe the presence of an optimal working point, where the system and the bath are in resonance and the energy transfer is maximum. Finally, for some $\Gamma \lesssim h_x$, the bath becomes too strong: the bath dynamics tends to temporarily trap the system into a steady state with respect to the instantaneous $h(t)$, which is typically not the final steady state, thus relaxation becomes slower for increasing $\Gamma$. From Fig.~\ref{fig:optimal} we also notice that, upon moving towards the critical line, the optimal working point moves to lower values of the coupling $\Gamma$. Moreover, we note that, albeit in our discussion we made use of values of the coupling $\Gamma$ of the order of $10^{-1}$ for better plot rendering, the same behavior is observed for lower values of the coupling, i.e., in the weak-coupling regime where the LE is, in general, more physically sound.

We verified the existence of an optimal working point also in a single-layer system with second-order phase transition like the one described by Eq.~\eqref{eq:generic-H} for $J_n=0$, $\forall \, n\neq 2$ (see also Ref.~\cite{nava2019lindblad}) as shown in the inset of Fig.~\ref{fig:time}. In such a way, we support the idea that the presence of the optimal working point is not a consequence of the inhomogeneities induced by the fixed boundary condition but, rather, an intrinsic property of the self-consistent dissipative dynamics.

We would like to emphasize that the model investigated here constitutes a simplified toy model for quantum wetting. Therefore, we do not yet aim to actively control $\tau(\Gamma)$ in experimental applications. Nevertheless, we do not exclude the possibility that this might be probed in experiments, involving, e.g., ultracold atoms~\cite{duan2003controlling,jepsen2020spin}, by tuning the system-bath interaction $\alpha$ (see Eq.~\eqref{eq:microscopic_H}).

The results shown in Fig.~\ref{fig:time} can be extended to the full phase diagrams as done for the wetting amplitude $\mathcal{A}$ suggesting that also the relaxation time can be used to extract, numerically or experimentally, the critical line in the coexistence region.

\section{Conclusions}
\label{sec:conclusion}

We have investigated the main static and dynamic features of the wetting interface within the coexistence region of a first-order transition, both quantum and thermal, when at the surface the metastable phase is favored over the stable one present in the interior of the bulk. For that, we have considered a prototypical mean-field model that displays a first-order phase transition both at zero and finite temperature, a fully-connected quantum Ising model with two and four spin exchange in slab geometry. Our work offers a new perspective on an old
topic, which is potentially wide though hindered by the simple mean-field toy model that we investigate. Such a new viewpoint could help to explore open questions on the quantum wetting transition and its critical properties \cite{quantum_wetting_1, quantum_wetting_2}. Indeed, instead of using a time-dependent Cahn's free-energy functional, as usually done in the literature, we have simulated the dynamics through the Lindblad master equation, where the temperature is directly inherited by the coupling to a dissipative bath. In this way, we were able to study the wetting phenomenon at any temperature and Hamiltonian parameters. In particular, we have reproduced the known critical behavior of the wetting interface length as the first-order transition is approached~\cite{lipowski1982critical}; also, we have identified a critical behavior of the relaxation time with bath-dependent exponents and the emergence in the parameter space of a dissipative optimal working point where the relaxation time is minimum.

Our analysis suggests a way to characterize the phase diagram alternative to the direct comparison between the free energies of the coexisting phases: our approach exploits the critical behavior in space and time of the wetting interface upon approaching the phase transition.  
The reliability of our approach in recovering physically sound results, combined with its simplicity and versatility, could make it a precious tool to investigate both equilibrium and non-equilibrium phase transitions in open quantum systems, paving the way to search for novel phases or phase transitions arising in spin models~\cite{tsvelik2013majorana,giuliano2020tunable} or junctions of interacting fermionic systems~\cite{oshikawa2006junctions,hou2008junctions,giuliano2020tunablespin,giuliano2015dual,buccheri2022violation,giuliano2022multiparticle,guerci2021probing}.

Moreover, to the best of our knowledge, detailed studies on the behavior of the relaxation time of the quantum wetting interface within dissipative dynamics have not been performed so far. In the future, it would be interesting to investigate the universal/non-universal behavior of the $\alpha$ critical exponent in both first- and second-order phase transitions, starting from simpler models without fixed boundary conditions. In addition, it would be desirable to find a microscopic (system+bath) Hamiltonian that gives rise to a Markovian master equation similar to the one employed in this work.

\section*{Acknowledgements}

This work received funding from the European Research Council (ERC) under the European Union’s Horizon 2020 research and innovation program, Grant agreements No.~101001902 (C.~A.) and No.~692670 (M.~F.). A.~N. acknowledges financial support from Italy’s MIUR PRIN projects TOP-SPIN (Grant No. PRIN 20177SL7HC).

\bibliography{bib}

\end{document}